\documentclass[aps,twocolumn,preprintnumbers,showpacs]{revtex4}
\usepackage{graphicx}
\usepackage{dcolumn}
\usepackage{bm}
\usepackage{CJK}
\usepackage{amsfonts}
\usepackage{psfrag}
\usepackage{wrapfig}
\usepackage{subfigure}
\usepackage{makeidx}
\usepackage{multirow}
\usepackage{epsf}
\usepackage{hyperref}
\usepackage{amsmath}
\usepackage{cases}

\begin{document}
\title{Stable Soliton Excitations in Modulational Instability Regime with the Fourth-order Effects}
\author{Liang Duan$^{1,2}$}
\author{Li-Chen Zhao$^{1,2}$\footnote{Electronic address: zhaolichen3@nwu.edu.cn}}
\author{Wen-Hao Xu$^{1,2}$}
\author{Chong Liu$^{1,2}$}
\author{Zhan-Ying Yang$^{1,2}$\footnote{Electronic address: zyyang@nwu.edu.cn}}
\author{Wen-Li Yang$^{2,3}$}

\address{$^{1}$School of Physics, Northwest University, 710069, Xi'an, China}
\address{$^{2}$Shaanxi Key Laboratory for Theoretical Physics Frontiers, 710069, Xi'an, China}
\address{$^{3}$Institute of Modern Physics, Northwest University, 710069, Xi¡¯an, China}


\begin{abstract}
We study the correspondence between modulational instability and types of fundamental nonlinear excitation in a nonlinear fiber with both third-order and fourth-order effects. Some stable soliton excitations are obtained in modulational instability regime, which have not been found in nonlinear fibers with second-order effects and third-order effects.  Explicit analysis suggests that the stable soliton existence is related with the modulation stability circle in the modulation instability regime, and they just exist in the modulational instability regime outside of the modulational stability circle. It should be emphasized that the stable soliton just exist with two special profiles on a continuous wave background with certain frequency.  The evolution stability of the solitons is tested numerically, which indicate they are robust against perturbations even in modulation instability regime. The further analysis indicates that the solitons in modulational instability regime are caused by the fourth-order effects.

\end{abstract}


\maketitle
\section{Introduction}
Modulation instability (MI) is a fundamental process associated
with the growth of perturbations on a continuous-wave(CW) background \cite{MI}.  It can be used to understand the dynamics of Akhmediev breather (AB) \cite{AB,Dudley}, Peregrine rogue wave (RW)\cite{RW,Kibler},  and Kuznetsov-Ma breather (K-M) \cite{K-M,Kibler2} and even high-order RWs\cite{He,Ling,ling1,Chabchoub,Chabchoub2,Chabchoub3}. Different MI gain distribution could bring different nonlinear excitation pattern dynamics \cite{zhao1}. We presented  rational W-shaped soliton in modulational stability (MS) regime \cite{zhao2}, and an autonomous transition dynamics from MI to MS regime on critical boundary lines between MI and MS regimes \cite{zhao3}. Recently few authors suggested that
baseband MI or MI with resonant perturbations could induce RW excitation, as a universal property of
different nonlinear models \cite{BasM,Basm,ChenF,Zhaoling}.  Furthermore, quantitative correspondence relations between nonlinear excitations and MI were clarified based on dominant perturbation frequency for the simplest nonlinear Schr\"{o}dinger equation (NLSE) \cite{Zhaoling}. The correspondence between MI and nonlinear excitations will be very meaningful for controllable excitation \cite{SR}. Therefore, it is meaningful and essential to obtain correspondence relation between MI and nonlinear excitations for other physical cases \cite{report}.

High-order effects are usually taken to describe the nonlinear dynamics more precisely for many different physical systems. For example, the nonlinear susceptibility will produce high-order nonlinear effects like the
Kerr dispersion (i.e., self-steepening) and the delayed nonlinear response, and even the third-order dispersion, for ultrashort pulses whose duration is shorter than 100fs in a nonlinear fiber. Recently, we obtained  correspondence relations between MI and nonlinear excitations for  Sasa-Satsuma (S-S) \cite{zhao3} and Hirota model \cite{Liu} which are the NLSE with some third-order effects. Those results indicate that the third-order effects brings great variation on MI property, which would induce some new nonlinear excitations. We extend the previous studies to consider the NLSE with both third-order and fourth-order effects (see Eq. (\ref{equ:1})), and perform linear stability analysis on a continuous wave background. We find an MS circle in the MI regime (see Fig. 1). This is topologically different from the finite width MS band for S-S model \cite{zhao3} and the MS line for Hirota equation\cite{Liu}. We have obtained many different nonlinear dynamics in the NLSE with high-order effects. For example, rational W-shaped soliton \cite{zhao1} and multi-peak soliton \cite{Liu} have been reported. It is naturally expected that there should be much more abundant nonlinear excitation dynamics in this case.

In this paper, we  study on the correspondence relation between MI and fundamental excitations in the NLSE with both third-order and fourth-order effects. These fundamental excitation mainly including  Akhmediev breather (AB), rogue wave (RW), Kuznetsov-Ma breather (K-M), periodic wave (PW), W-shaped soliton train (WST), rational W-shaped soliton (WS$_{r}$), anti-dark soliton (AD) and non-rational W-shaped soliton (WS$_{nr}$), can be described by an exact solution in a generic form.  Especially, stable AD and WS$_{nr}$ soliton excitations can exist in MI regime outside of the MS circle, in sharp contrast to the correspondence relation for the simplest NLSE \cite{Zhaoling}, Hirota equation \cite{Liu}, and S-S equation \cite{zhao3}. When the soliton perturbation energy on CW background tend to zero, the stable perturbation agrees well with the stability property predicted by linear stability analysis. But linear stability analysis fails to explain the stability of soliton with larger perturbation energy on the CW background, since the linear stability analysis does not hold anymore for the large amplitude perturbations on CW background. It should be emphasized that the stable soliton just exist with a special profile on a certain frequency continuous wave background, and the stable soliton existence in MI regime is caused by the fourth-order effects. We further test the evolution stability of the stable solitons numerically, which indicate they are robust against perturbation even in MI regime.

Our presentation of the above features will be structured
as follows. In Sec. II,  the linear analysis on a CW background is performed and a generic exact solution is given for types of fundamental nonlinear excitations. The correspondence between MI and these fundamental nonlinear excitations is presented. Especially, stable anti-dark soliton and non-rational W-shaped soliton can exist in MI regime, which is absent for NLSE, S-S, and Hirota model.  In
Sec. III, we discuss the anti-dark soliton and non-rational W-shaped soliton in MI regime and test the stability of them numerically.  It is found that the stable soliton just exist with a special profile on a certain frequency plane wave background, and the stable soliton existence in MI regime is caused by the fourth-order effects.  Finally,  we summarize the results and
present our discussions in Sec. IV.

\section{The correspondence between modulational instability and several fundamental nonlinear excitations}
Many efforts have been paid to study nonlinear excitations in nonlinear fiber with high-order effects, such as  S-S equation \cite{zhao2,zhao3,Zhaoling2, SS,TaoXu, TaoXu2}, and Hirota equation \cite{Hirota,Hirota2,Liuzhao}. The S-S and Hirota equations are the NLSE with third-order effects. Some recent experiments suggested that fourth-order effects could play important role on soliton dynamics in nonlinear fiber \cite{pfs}.  We would like to further consider a NLSE with both third-order and  forth-order effects as follows \cite{Adrian, Chowdury, Yunqing, Ankewwicz, Wang}
\begin{equation}
\label{equ:1}
i\psi_z+\frac{1}{2}\psi_{tt}+\psi|\psi|^2+i\beta H[\psi(t,z)]+\gamma P[\psi(t,z)]=0
\end{equation}
where the third-order $H[\psi(t,z)]=\psi_{ttt}+6|\psi|^2\psi_t$ is the Hirota operator (beginning with third-order dispersion),
and the fourth-order $P[\psi(t,z)]=\psi_{tttt}+8|\psi|^2\psi_{tt}+6\psi|\psi|^4\nonumber\\
+4\psi|\psi_t|^2+6\psi_t^2\psi^*+2\psi^2\psi_{tt}^*$ is the Lakshmanan-Porsezian-Daniel (LPD) operator (beginning with fourth-order dispersion).
Here, $z$ is the propagation variable and $t$ is the transverse variable (time in a moving frame), with the function $|\psi(t,z)|$ being the envelope of the waves.

Since MI properties can be used to understand different excitation patterns in nonlinear systems \cite{Zhaoling,zhao2,zhao3,Liu}, we perform the standard linear stability analysis on a generic continuous wave (CW) background $ \psi_{0}=Ae^{i\theta}=A\  e^{i (k z+\omega t )}$ ($A$, $\omega$ and $k=A^{2}-\frac{1}{2}\omega^{2}+\beta(\omega^{3}-6 A^{2}\omega)+\gamma(6A^{4}-12A^{2}\omega^{2}+\omega^{4})$ represent the amplitude and frequency and wave number of background electric field, respectively). A perturbed nonlinear background can be written as $\psi_{p}=(A+p)e^{i\theta}$, where $p(t,z)$ is a small perturbation which is given by collecting the Fourier modes as $p=f_{+}\ exp[i(Kz+\Omega t)]+f_{-} \ exp[-i(Kz+\Omega t)]$, where $f_{+}, f_{-}$ are much less than the background amplitude $A$, $\Omega$ represents perturbed frequency. Then, one can obtain the modulation instability (MI) gain $G=| Im(K)|= Im{\Omega \sqrt{(\Omega^{2}-4A^{2})[1/2-3\beta\omega+\gamma(6A^{2}-6\omega^{2}-\Omega^{2})]^{2}}}$ through linearizing the nonlinear partial equation and solving the eigenvalue problems. Obversely, there a low perturbed frequency MI regime ($|\Omega|<2A$) in perturbed frequency $\Omega$ and background frequency $\omega$ space.

\begin{figure}[htb]
\centering
{\includegraphics[height=105mm,width=70mm]{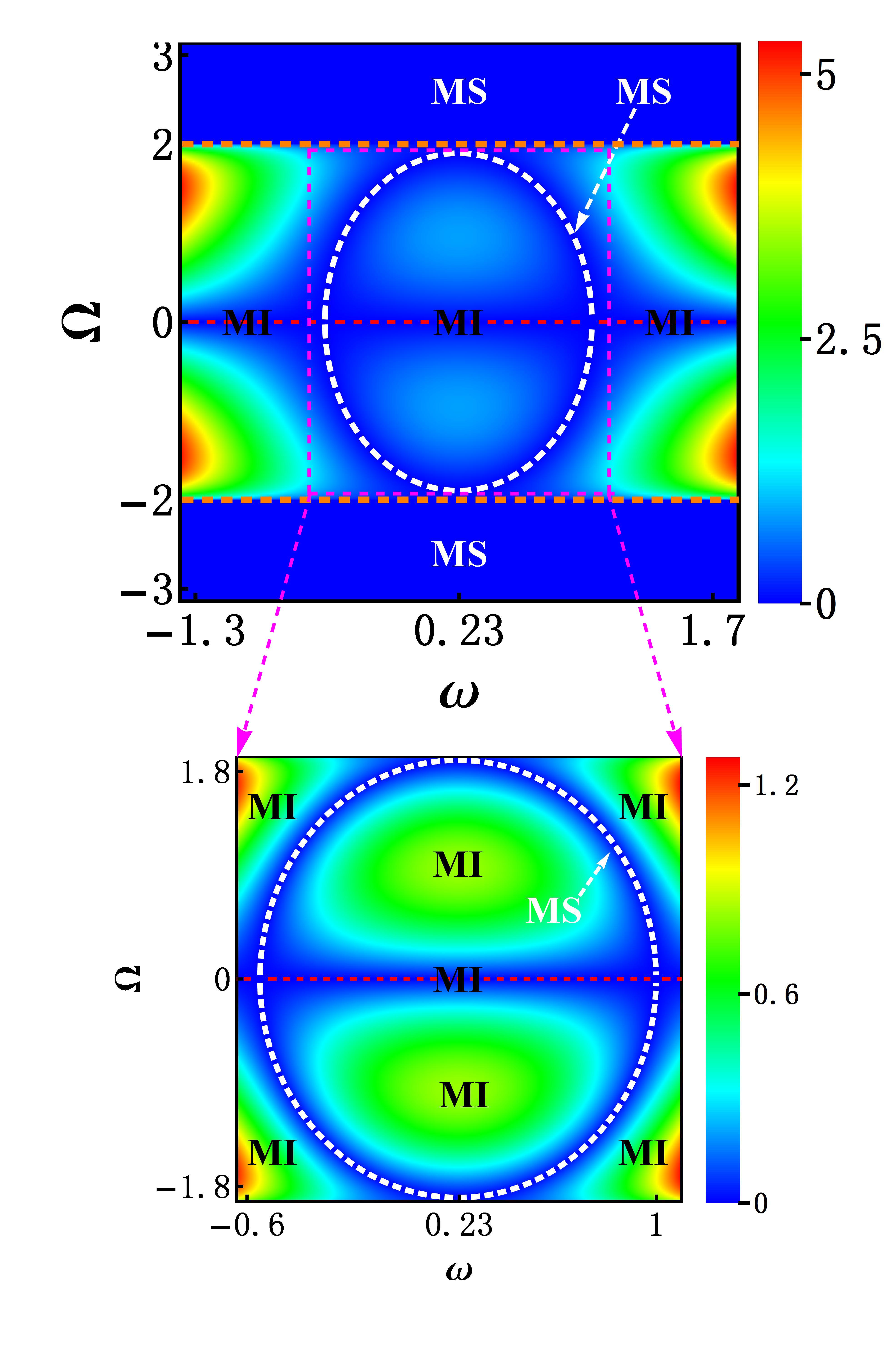}}
\caption{ Modulation instability gain $G$ distribution on background frequency $\omega$ and perturbation frequency $\Omega$ plane, where MI and MS denote modulation instability and modulation stability. There is a MS circle in the MI band, marked by a white dashed line. The MI gain value is not zero for the regime outside and inside of the MS circle, shown in the lower frame. The parameters are $A=1$, $\beta=\frac{1}{6}$ and $\gamma=-\frac{5+\sqrt{15}}{48}$.}
\end{figure}
Particularly, when $1/2-3\beta\omega+\gamma(6A^{2}-6\omega^{2}-\Omega^{2})=0$, i.e.
\begin{eqnarray}
\label{equ:7}
(\omega+\frac{\beta}{4\gamma})^{2}+\frac{\Omega^{2}}{6}=\alpha,
\end{eqnarray}
the MI gain $G=0$, where $\alpha=\frac{\beta^{2}}{16\gamma^{2}}+\frac{1}{12\gamma}+A^2$.
When $\alpha>0$, the MS regime Eq. (\ref{equ:7}) describes an ellipse whose center is localized in point ($-\frac{\beta}{4\gamma}$, $0$). Its semi-major axis is equal to $\sqrt{6\alpha}$ and parallel to the $\Omega$ coordinate and semi-minor axis is equal to $\sqrt{\alpha}$ in $\omega$ coordinate.  When semi-major axis less than $2 A$ and greater than $0$, all parts of the ellipse are located in the MI regime [See Fig. 1].  Namely,
the MI band ($|\Omega| \leq 2A$) contain special MS regime which
satisfy Eq. (2), which brings MS circle in the MI regime. Nonzero MI gain value exist on both outside and inside areas of the circle [see Fig. 1]. It should be noted that the MI gain form fails to predict the stability of perturbations on the resonant line ($\Omega=0$) \cite{Zhaoling}. For $\Omega=0$ mode perturbation denoted by  $\epsilon \tilde{p}$ (where $\epsilon\ll1$ is a real constant), we can derive the secular solution as  $\tilde{p}=1+24iA^{2}\gamma\left[\alpha-\left(\omega+\frac{\beta}{4\gamma}\right)^{2}\right]z$ which demonstrates the instability property of the resonant perturbation mode. The perturbations with modes $\left(\omega+\frac{\beta}{4\gamma}\right)^{2}=\alpha$ on the MS circle do not admit rational growth.

This is topologically different from the MI distribution for the
simplest NLSE, Hirota, and S-S equations \cite{Zhaoling,Liu,zhao3}.
We expect that there could be some exotic dynamical excitations for this new MI distribution pattern. Moreover, the MS circle size can be changed by varying the fourth-order effect strength. When semi-major axis greater than $2A$, only a part of the ellipse is located in the MI band ($|\Omega|\leq2A$). In this case, MS regime which is located in the MI band are two curves. Moreover, when $\alpha=0$, the MS circle reduce to be a MS point ($-\frac{\beta}{4\gamma}$, $0$) which is located on the resonance line in MI band.  Furthermore, for $\alpha<0$, there is no MS regime in the MI band. To show the correspondence between MI and nonlinear excitations conveniently and clearly, we choose Fig. 1 to study on the relation. Similar discussion can be made for other cases, and we have proved that the nonlinear excitations are the most abundant for the case in Fig. 1.

To obtain the correspondence relation between fundamental nonlinear waves and MI distribution, we turn our attention to the exact solutions on the continuous wave (CW) background. We derive an exact solution by means of the Darboux transformation method, which can describe many different types of fundamental nonlinear excitations. The explicit expressions for the solution are presented in the Appendix. The solution (\ref{equ:A1}) is a nonlinear combination of trigonometric function ($\cos\varphi$) and hyperbolic function ($\cosh\phi$), where $\varphi$ and $\phi$ are real functions of $z$ and $t$ [see Eq. (\ref{equ:A2})]. Here the hyperbolic function and trigonometric function describe the localization and the periodicity of the nonlinear waves, respectively. Hence this nonlinear wave described by the solution (\ref{equ:A1}) can be seen as a nonlinear superposition of a soliton and a periodic wave. We find that the solution (\ref{equ:A1}) contain eight types fundamental nonlinear excitations, such as Kuznetsov-Ma breather (K-M), non-rational W-shaped soliton (WS$_{nr}$), anti-dark soliton (AD), Akhmediev breather (AB), periodic wave (PW), W-shaped soliton train (WST), rogue wave (RW), and rational W-shaped soliton train (WS$_{r}$). The existence conditions and explicit expressions of these fundamental nonlinear excitations are given in Table I.
\begin{widetext}
\makeatletter
\newcommand{\tabcaption}{\def\@captype{table}\caption}
\makeatother
\renewcommand\arraystretch{1.8}
\begin{center}
\footnotesize
 \label{table1}
\begin{tabular}{|c|c|c|}
 \hline
  Existence condition & Nonlinear waves type  & Analytic expression \\
  \hline
  $|b|>A$, $(\omega+\frac{\beta}{4\gamma})^{2}-\frac{\delta^{2}}{6}\neq\alpha$ & Kuznetsov-Ma breather & $\left[1-\frac{2b[\Lambda\cos(bv_{2}\delta z-\theta_{k})-\cosh(\chi_{k})]}
{A\cos(bv_{2}\delta z)-b\cosh(\chi_{k})}\right]Ae^{i\theta}$\\
\hline
  $|b|<A$, $(\omega+\frac{\beta}{4\gamma})^{2}+\frac{\Omega^{2}}{6}\neq\alpha$ & Akhmediev breather & $\left[-A-\frac{(2A^{2}-2b^{2})\cosh(bv_{2}\Omega z)+i\Omega b\sinh(bv_{2}\Omega z)}{b\cos(\chi_{a})- A\cosh(2bv_{2}\Omega z)}\right]e^{i\theta}$ \\
\hline
  $|b|=A$, $(\omega+\frac{\beta}{4\gamma})^{2}\neq\alpha$ & rogue wave & $\left[\frac{4(2iA^{2}v_{2}z+1)}{1+4A^{2}(t+v_{r}z)^{2}+4A^{4}v_{2}^{2}z^{2}}-1\right]Ae^{i\theta}$ \\
  \hline
  $|b|>A$, $(\omega+\frac{\beta}{4\gamma})^{2}-\frac{\delta^{2}}{6}=\alpha$ & non-rational W-shaped soliton & $\left[\frac{-\delta^{2}}{2A-2b\cosh\chi_{s}}-A\right]e^{i\theta}$\\
  \hline
   $|b|>A$, $(\omega+\frac{\beta}{4\gamma})^{2}-\frac{\delta^{2}}{6}=\alpha$ & anti-dark soliton & $\left[\frac{-\delta^{2}}{2A+2b\cosh(\chi_{s}-\mu_{2})}-A\right]e^{i\theta}$\\
  \hline
  $|b|=A$, $(\omega+\frac{\beta}{4\gamma})=\alpha$, $\alpha\ge0$ & rational W-shaped soliton & $[\frac{4}{1+4A^{2}(t+v_{w}z)^{2}}-1]Ae^{i\theta}$ \\
  \hline
  $\frac{A}{2}<|b|<A$, $(\omega+\frac{\beta}{4\gamma})^{2}+\frac{\Omega^{2}}{6}=\alpha$, $\alpha>\frac{A}{2}$ & W-shaped soliton train & $[\frac{\Omega^{2}}{2A-2b\cos\chi_{p}}-A]e^{i\theta}$ \\
  \hline
  $0<|b|\le\frac{A}{2}$, $(\omega+\frac{\beta}{4\gamma})^{2}+\frac{\Omega^{2}}{6}=\alpha$, $\alpha>0$ & periodic wave & $[\frac{\Omega^{2}}{2A-2b\cos\chi_{p}}-A]e^{i\theta}$ \\
  \hline
\end{tabular}
\tabcaption{Types of nonlinear waves in forth-order NLS system. The parameters are $\delta=2\sqrt{b^{2}-A^{2}}$, $\Omega=\pm2\sqrt{A^{2}-b^{2}}$, $\chi_{k}=\delta(t+v_{k}z)$, $\chi_{a}=\Omega(t+v_{a}z)$, $\chi_{s}=\delta(t+v_{s}z)$, $\chi_{p}=\Omega(t+v_{p}z)$, $v_{k}$, $v_{a}$, $v_{r}$, $v_{s}$, $v_{w}$, and $v_{p}$ correspond to the $v_{1}$ in Appendix in corresponding conditions, $\Lambda=\frac{\sqrt{s^{2}}}{s}$($s=b+\sqrt{b^{2}-A^{2}}$, $|b|>A$), and the phase satisfy $\cos\theta_{k}=\frac{bs}{A\sqrt{s^{2}}}$, $\sin\theta_{k}=i\frac{bs-A^{2}}{A\sqrt{s^{2}}}$, $\cosh\mu_{AD}=b/A$, $\sinh\mu_{AD}=\sqrt{b^{2}-A^{2}}/A$, and the other parameters are the same as those in Appendix.}
\end{center}
\end{widetext}

\begin{figure}[htb]
\centering
{\includegraphics[height=36mm,width=39mm]{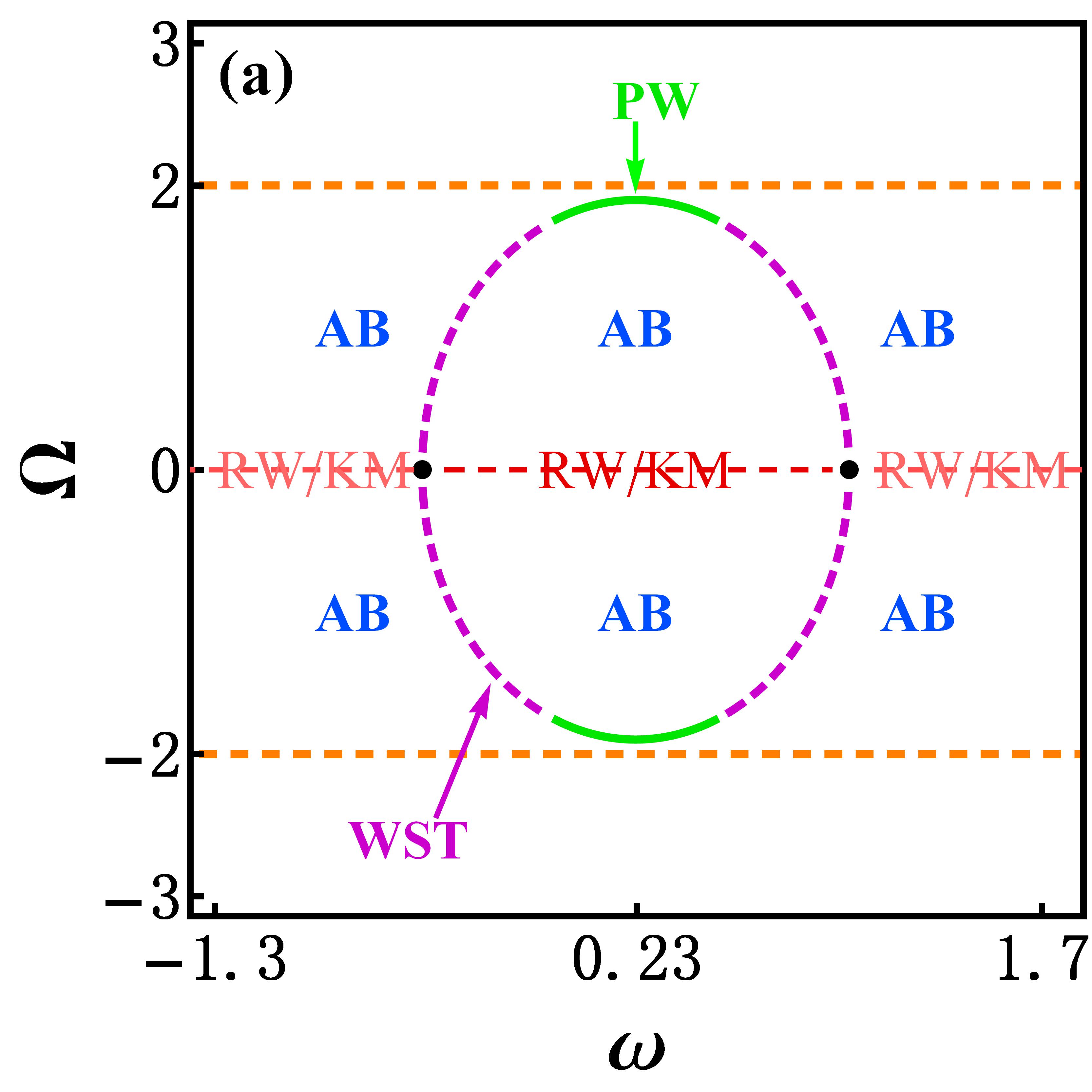}}
{\includegraphics[height=36mm,width=39mm]{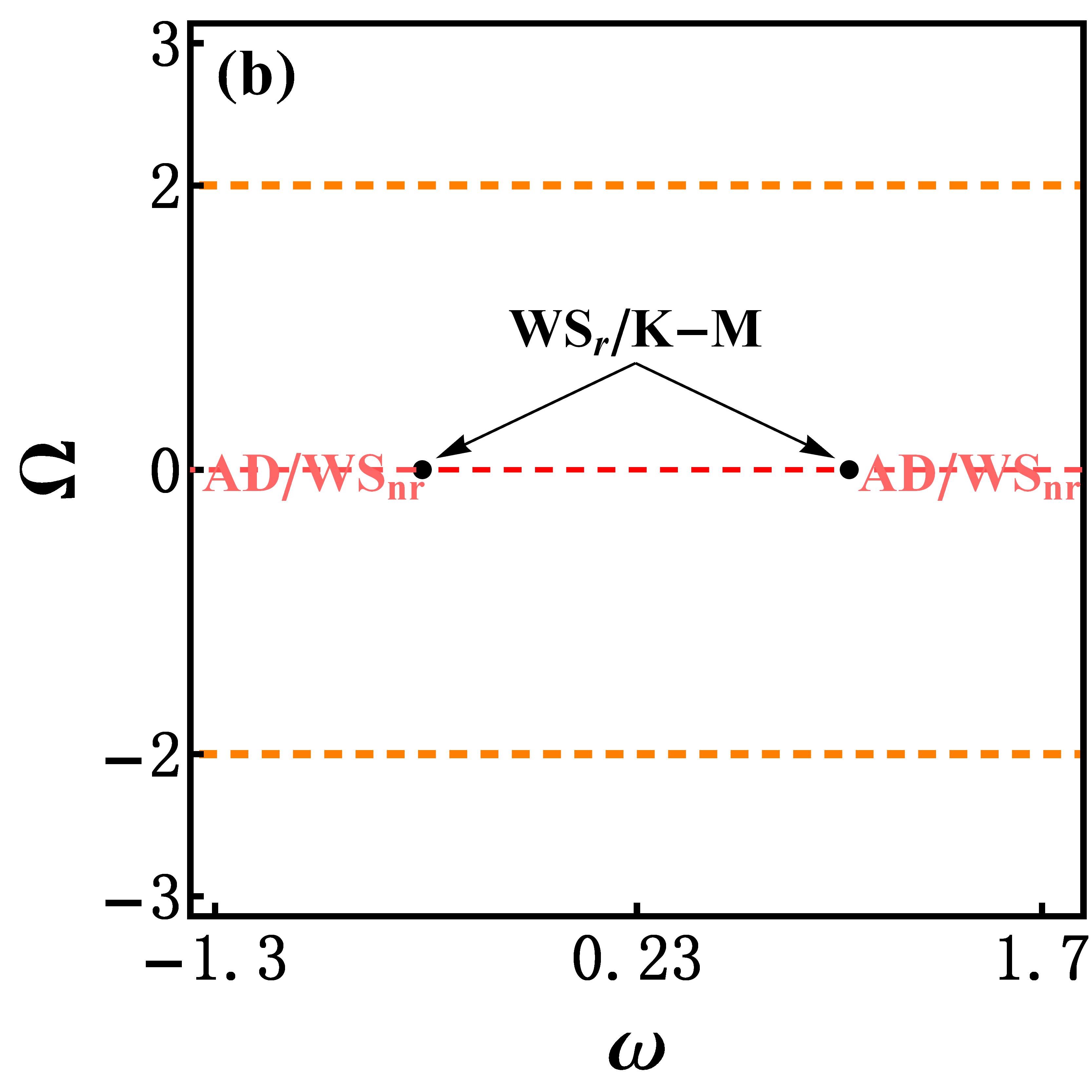}}
\caption{(a) Phase diagram for different types nonlinear waves on modulation instability gain spectrum plane corresponding to Fig. 1. ``AB", ``RW", ``K-M", ``PW", ``WST", ``WS$_{r}$", ``AD", and ``WS$_{nr}$"denote  Akhmediev breather, rogue wave, Kuznetsov-Ma breather, periodic wave, W-shaped soliton train, rational W-shaped soliton, anti-dark soliton and non-rational W-shaped soliton respectively. Especially, the stable AD and WS$_{nr}$ can exist on the resonant line in MI regime shown in (b). The parameters are the same with Fig. 1.}
\end{figure}

Based on Fourier analysis of the exact solutions of RW, AB, K-M, WS$_{r}$, WS$_{nr}$, WST, AD, and PW, we locate these different fundamental nonlinear excitations on the MI plane through defining and calculating the dominant frequency and propagation constant of each nonlinear wave \cite{Zhaoling}, shown in Fig. 2(a) and (b).
We can see that the RW and K-M still comes from the resonance perturbation in MI regimes and the AB still in the regime between the resonance line and the boundary line between the MI and MS regime. These are similar to the RW, AB and KM for the NLSE system. However, there are MS lines in MI band for forth-order NLSE system and the PW, WST and WS$_{r}$ are excited in this regime [see Fig. 2(a)]. We have shown that the MS regime structures in MI regime depend on the parameter $\alpha$ [see Eq. (2)]. Therefore, stable nonlinear waves types are different as difference of parameter $\alpha$. When the semi-major axis $\sqrt{6\alpha}$ of ellipse (\ref{equ:7}) less than $\sqrt{3}A$ (i.e. $\alpha<\frac{A^{2}}{2}$), the PW would not exist. Moreover, when $\alpha=0$, the PW and WST do not exist. Furthermore, for $\alpha<0$, the stable nonlinear waves (PW, WST, and WS$_{r}$) on MS regime in MI band do not exist. These results are consistent with previous investigations. The MI regime corresponds to the instability nonlinear waves (i.e., breathers and RW) and MS regime localized in MI band corresponds to the excitations of the stable nonlinear waves (i.e., solitons and periodic wave). However, we find  AD and WS$_{nr}$ exist in the MI regime [see Fig. 2(b)], which seems violate with the prediction given by linear stability analysis. Particularly, they just exist in the MI regime outside of the MS circle, but they can not exist in the MI regime inside of the MS circle. We discuss them in detail in the next section.
\begin{figure}[htb]
\centering
\label{fig:3}
{\includegraphics[height=45mm,width=41mm]{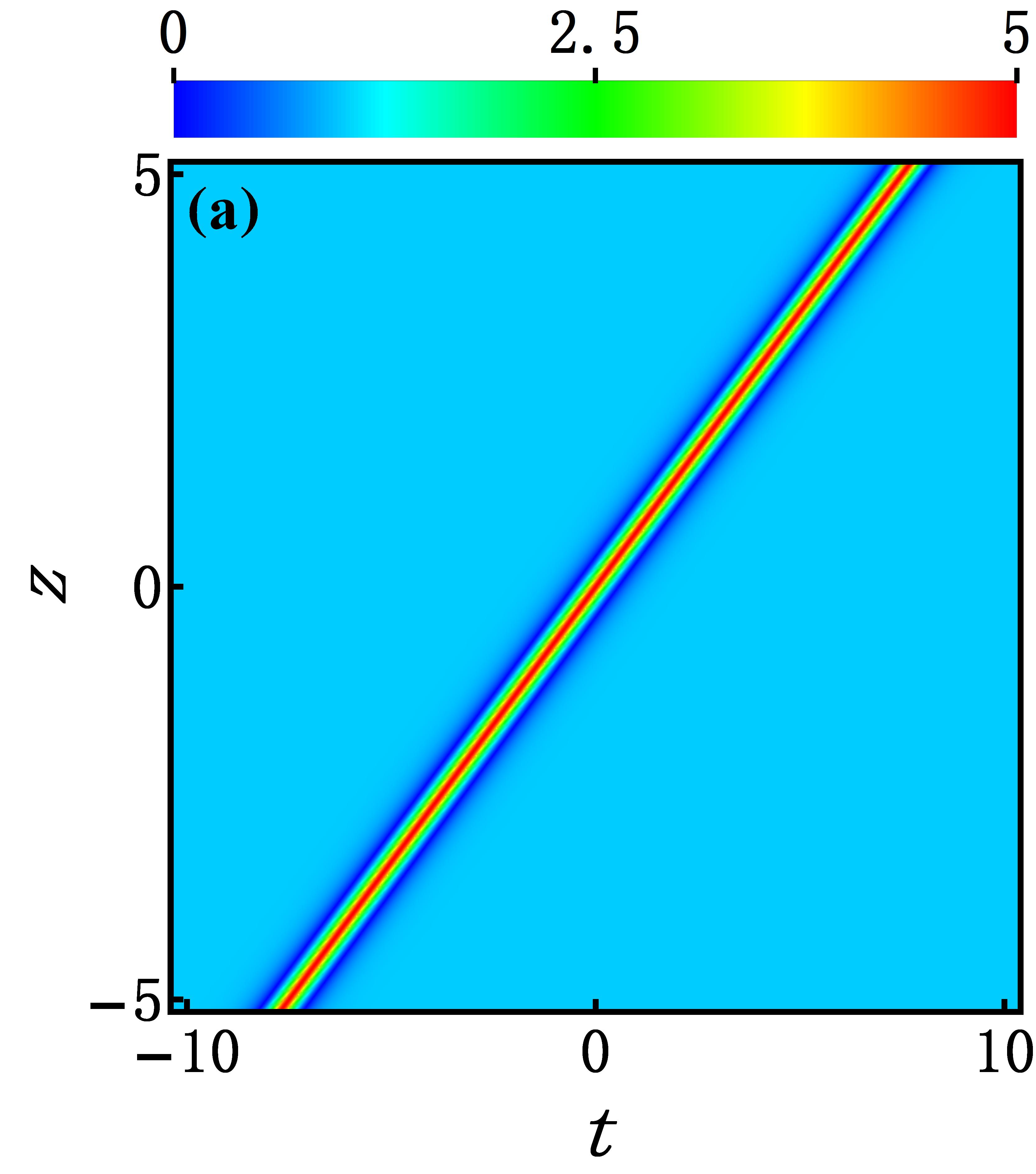}}\label{fig:3a}
{\includegraphics[height=45mm,width=41mm]{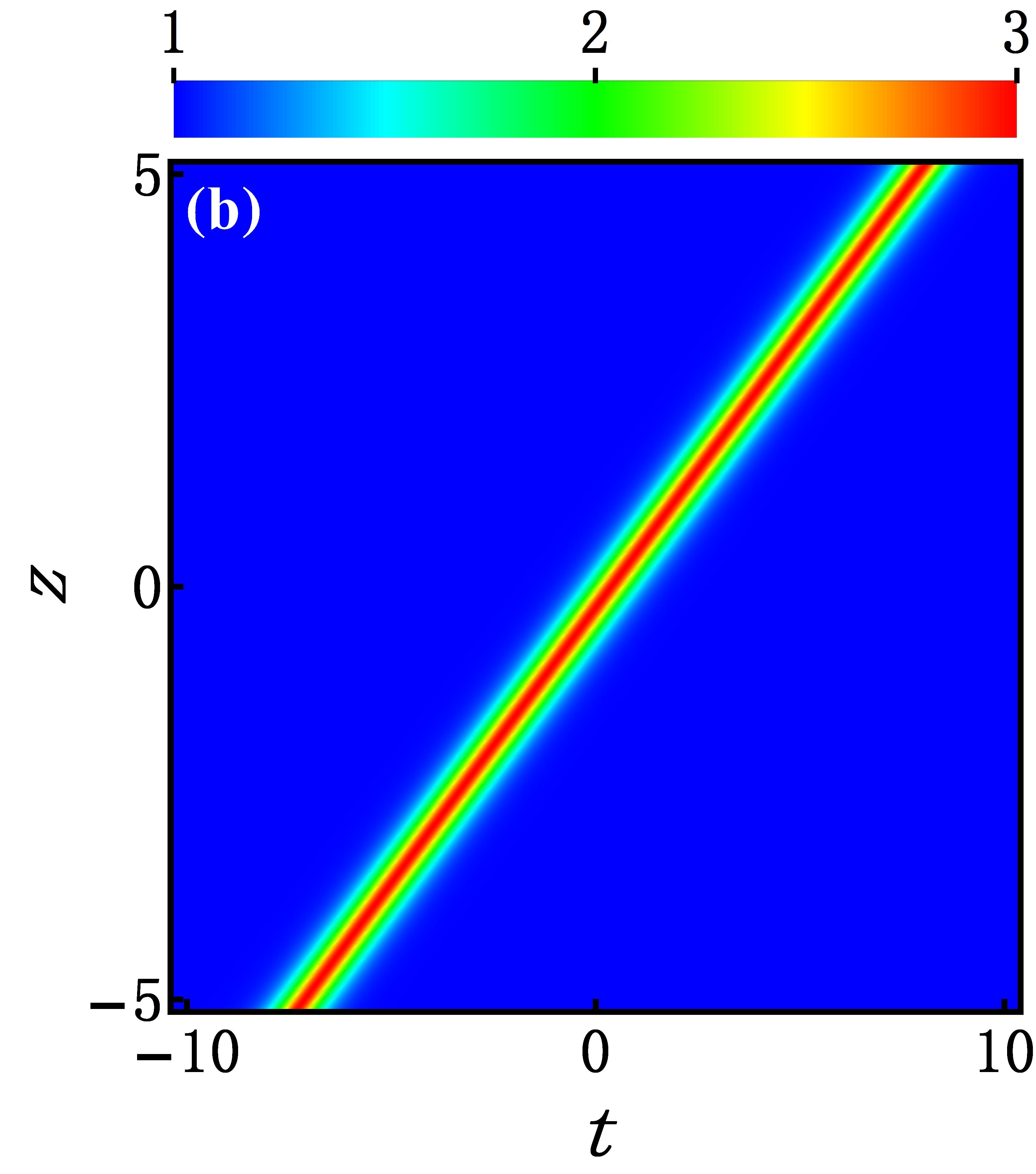}}\label{fig:3b}
\caption{(color online) Optical amplitude distributions $|\psi_{s}|$ of (a) the non-rational W-shaped soliton and (b) anti-dark soliton on a continuous wave background. The parameters are $A=1$, $\beta=\frac{1}{12}$, $\gamma=-\frac{1}{36}$, $\omega=0$, $b=2$.}
\end{figure}
\begin{figure}[htb]
\centering
{\includegraphics[height=45mm,width=41mm]{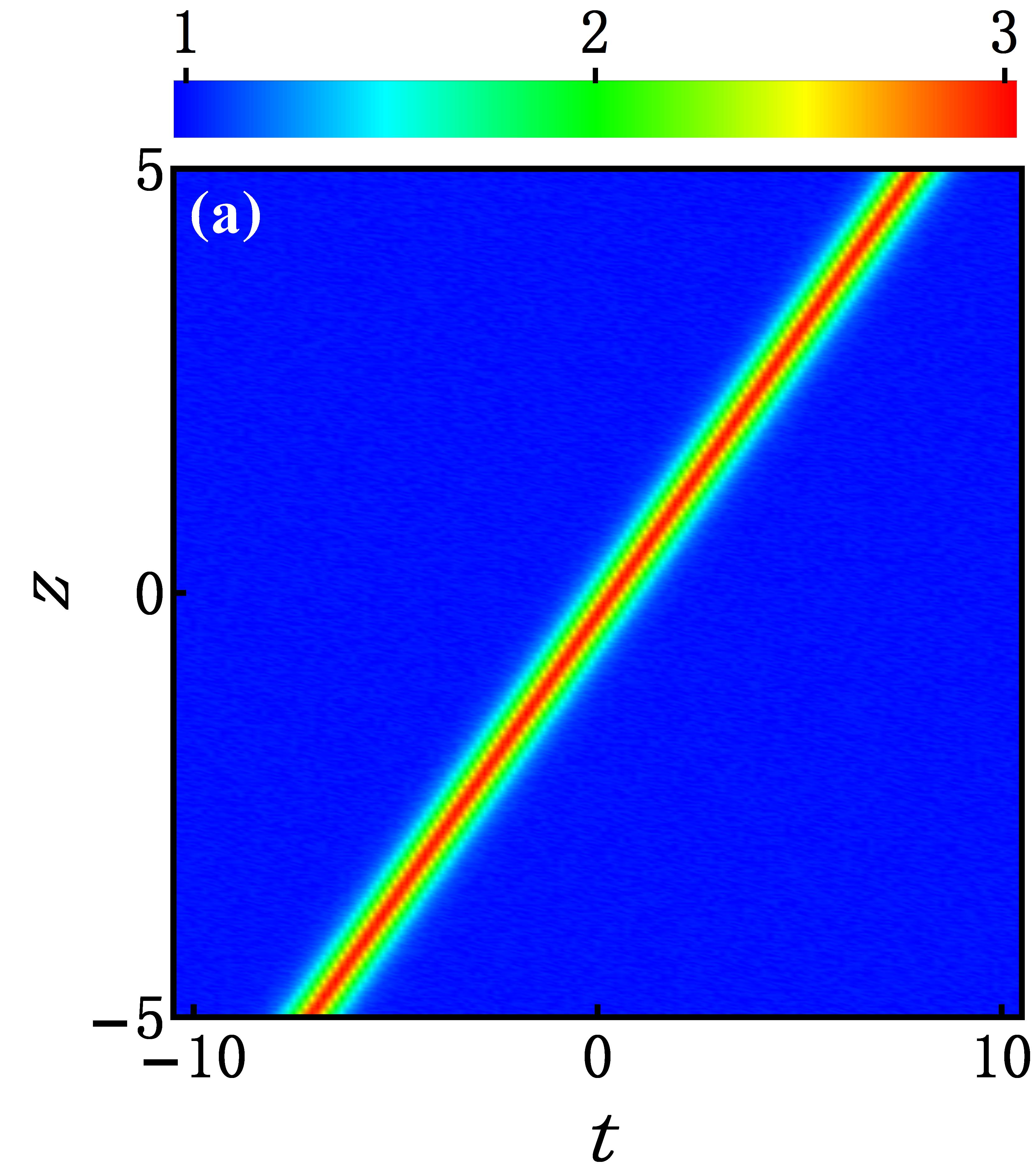}}
{\includegraphics[height=45mm,width=41mm]{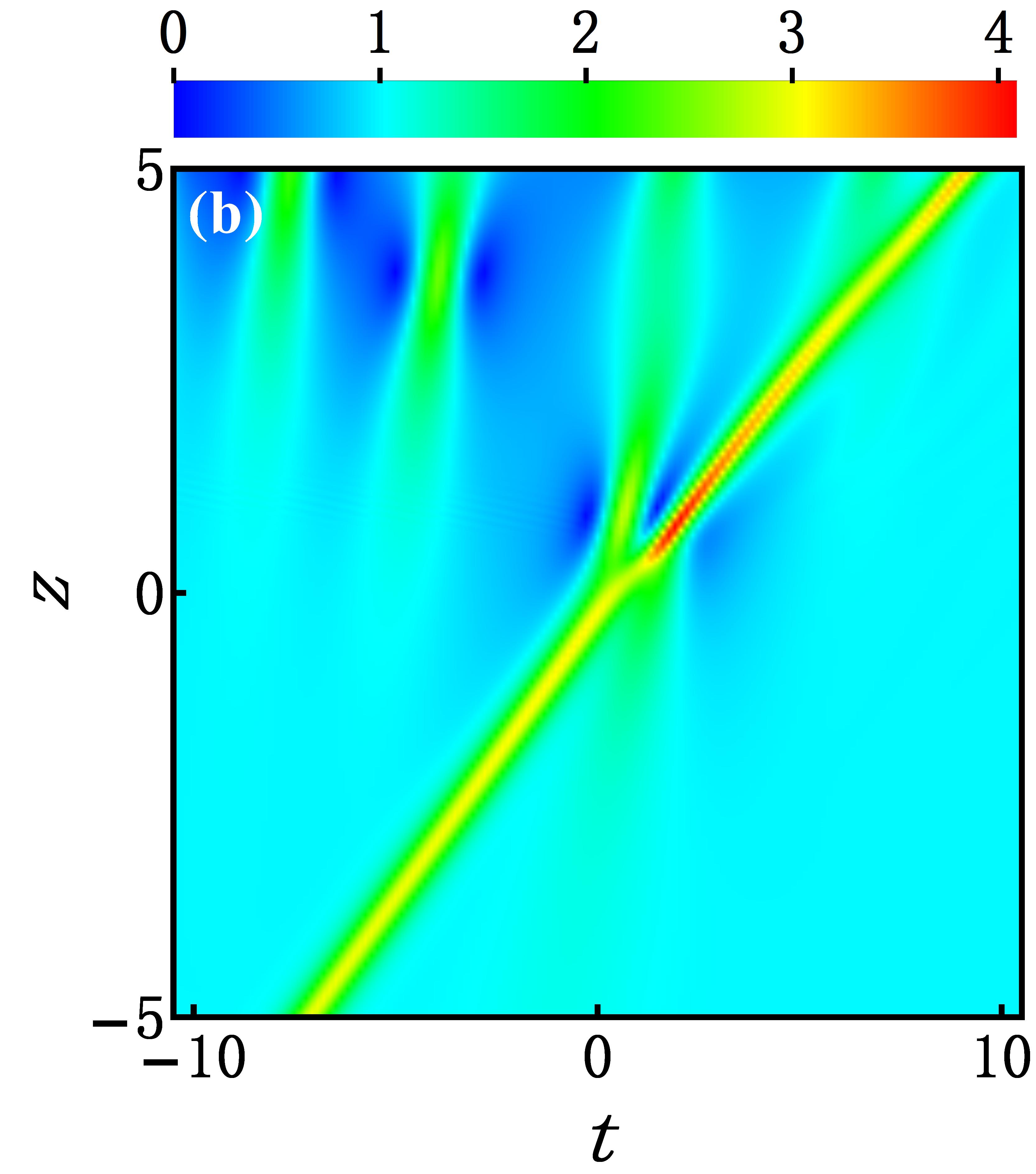}}
\caption{(color online) (a) Numerical evolution is from the initial conditions that $\psi_{p}(t,-5)=\psi_{AD}[1+0.01Random(t)]$. (b) Numerical evolution of $|\psi(t,z)|$ from the AD with adding a weak Gaussian pulse perturbations in the form of $\psi_{AD}[1+0.1e^{-(t+1)^{2}/4}]$. The parameters are $A=1$, $\beta=\frac{1}{12}$, $\gamma=-\frac{1}{36}$, $\omega=0$, $b=2$.}
\end{figure}

\section{Anti-dark soliton and W-shaped soliton in the MI regime}\label{sec:3}
The evolutions of AD and WS$_{nr}$  are shown in Fig. 3, which indeed evolve stably with invariable profiles. To confirm that these two types soliton can propagate stably, as an example, we numerically test the stability of AD against noises (the initial condition $\psi_{p}=\psi_{AD}[1+0.01Random(t)]$). The result is shown in Fig. 4(a), which indicate AD is robust against noises. Furthermore, we evolve AD with adding a weak Gaussian pulse perturbation $\psi_{AD}[1+0.1e^{-(t+1)^{2}/4}]$,  shown in Fig. 4(b). We can see that the AD propagate stably, and a weak Gaussian pule evolves into a RW. After interacting with the RW, the AD restore the original shapes with a small phase shift. This confirms that the AD indeed exist in the MI regime.  This is significantly different from previous results for which there is no stable localized wave in the MI regime \cite{Zhaoling,Liu,zhao2,zhao3}.
The above analysis shows that the AD and WS$_{nr}$ exist with the same parameter condition $(\omega+\frac{\beta}{4\gamma})^{2}-\frac{\delta^{2}}{6}=\alpha$ (where $\delta=2 \sqrt{b^2-A^2}$). The structures of these two kinds of solitons are different [see Fig. 3(a) and (b)], but the energy of the AD and WS$_{nr}$  against the CW background are identical, namely, $\varepsilon_{s1}=\varepsilon_{s2}=2\delta$. The perturbation energy $\varepsilon_{s}$ of pulse against the background is defined as
\begin{equation}
\varepsilon_{s}=\int_{-\infty}^{\infty}(|\psi|^{2}-A^{2})dt.
\end{equation}
Because there is a one-to-one correspondence between the parameter $\delta$ and the perturbation energy $\varepsilon$, we can rewrite the existence condition for AD and WS$_{nr}$ as follows
\begin{equation}
\label{equ:16}
(\omega+\frac{\beta}{4\gamma})^{2}-\frac{\varepsilon_{s}^{2}}{24}=\alpha.
\end{equation}
The condition has a more clear meaning since the perturbation energy has more clear physical meaning than the parameter $\delta$. Namely, for specific system parameter and fixed background amplitude (i.e. $\beta$, $\gamma$, $A$ are fixed), when background frequency $\omega$ and the energy of pulse $\varepsilon_{s}$ satisfy the Eq.(\ref{equ:16}), a CW background admits solitons with two special profiles (AD and WS$_{nr}$) even in MI regime (outside the MS circle). This character has not been found for the NLSE and its extended form with other types of high-order effects \cite{zhao2,zhao3,Zhaoling2, SS,TaoXu,TaoXu2,He,Hirota,Hirota2,Liuzhao,K-E,WangX}.

In order to better understand the generation mechanism of AD and WS$_{nr}$, we analyze the relationship between their existence conditions and MI gain. From the above linear stability analysis, the MS condition on resonance line can be given
$\omega=-\frac{\beta}{4\gamma}\pm\sqrt{\alpha}$.
This means that there are two MS points on the resonance line [see Fig. 2(a)]. Then we rewrite the existence conditions of AD and WS$_{nr}$ as follows
\begin{equation}
\label{equ:18}
\omega_{s}=-\frac{\beta}{4\gamma}\pm\sqrt{\frac{\varepsilon_{s}^{2}}{24}+\alpha}.
\end{equation}
It is seen that the AD and WS$_{nr}$ exist on both sides of two MS points on resonance line [see Fig. 2(b)](but do not exist in the regime between the two MS points) and their excitation positions depend on the pulse energy $\varepsilon_{s}$. Then we plot the positions of the AD and WS$_{nr}$ with different energies in Fig. 5. When the energy of the AD and WS$_{nr}$ approach zero (their amplitudes are small), the excitation condition will tend to MS points, which agrees well with the prediction of linear stability analysis. But  their excitation positions gradually deviate from the MS points as the energy increase. Namely, the energies of the AD and WS$_{nr}$ are non-zero, their excitation conditions do not agree with the MS condition predicted by linear stability analysis. Therefore, the stable soliton with high perturbation energy could exist in MI regime predicted by linear stability analysis. This means that perturbation energy could inhibit the MI property. Considering that stable soliton just can exist with a specific perturbation energy  on a certain CW background, we expect that there is a balance condition between perturbation energy and MI gain.

\begin{figure}[htb!]
\centering
{\includegraphics[height=64mm,width=75mm]{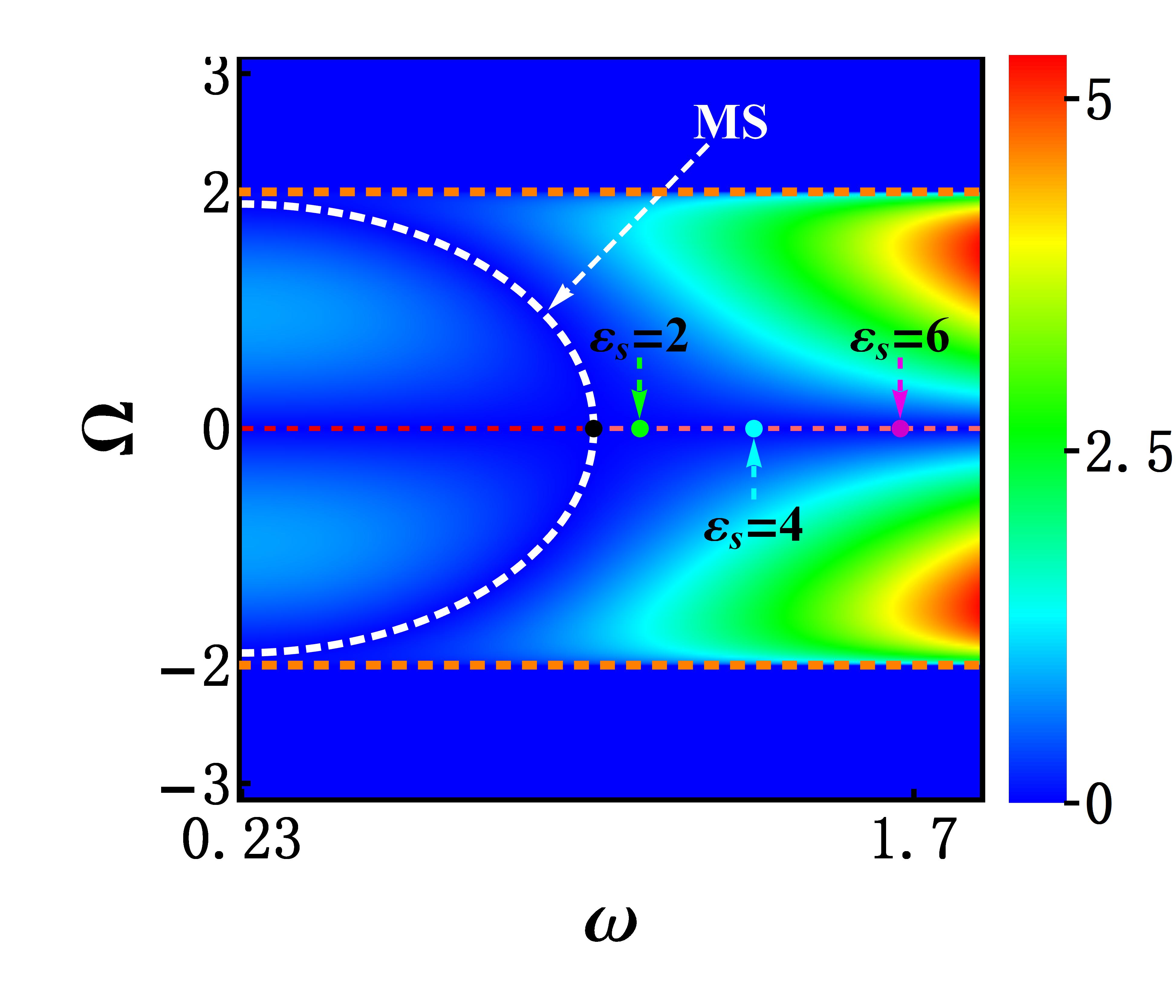}}
\caption{The positions of the AD and WS$_{nr}$ with different energies on the modulational instability gain spectrum plane.   Green, blue-green and purple points correspond to the AD and WS$_{nr}$ with different energy $\varepsilon_{s}$ and blue points is modulation stability points on the resonance line. It is shown that the position  approach to the MS point as the perturbation energy decrease (the soliton amplitude becomes smaller). The parameters are $A=1$, $\beta=\frac{1}{6}$, and $\gamma=-\frac{5+\sqrt{15}}{48}$.}
\end{figure}

Then, we would like to find out which factors bring the stable soliton in MI regime. The above analysis indicates that the MI gain $G$ and pulse energy $\varepsilon_{s}$ both play an important role on the excitation of the AD and WS$_{nr}$. To further analyze the impact of the MI gain $G$ and energy $\varepsilon_{s}$ on excitation of the AD and WS$_{nr}$, we simplify the expression of the gain $G$ on resonance line based on the above linear stability analysis result,
$G=24A^{2}|\gamma|\left|(\omega+\frac{\beta}{4\gamma})^{2}-\alpha\right|$.
On the other hand, we can rewrite the energy squared expression of the AD and WS$_{nr}$ from the Eq. (\ref{equ:16}), namely,
$\varepsilon_{s}^{2}=24\left[(\omega+\frac{\beta}{4\gamma})^{2}-\alpha\right]$.
The AD and WS$_{nr}$ do not exist when $(\omega+\frac{\beta}{4\gamma})^{2}-\alpha\leq0$ [see the part of between two MS points in Fig. 2(a) and (b)]. Therefore, we only consider the condition of $(\omega+\frac{\beta}{4\gamma})^{2}-\alpha>0$. In this case, MI gain $G$ and the energy of the AD and WS$_{nr}$ satisfy
\begin{equation}
\label{equ:21}
\frac{G}{\varepsilon_{s}^{2}}=A^{2}|\gamma|.
\end{equation}
This can be seen as a balance condition between perturbation energy and MI. Obviously,  the fourth-order effects related with $\gamma$ here bring the existence for AD and WS$_{nr}$ in MI regime.  This can be used to explain that the AD and WS$_{nr}$ in MI regime have not been found in  the NLSE and its extended form with lower than fourth-order effects \cite{TaoXu,TaoXu2,He,Hirota,Hirota2,Liuzhao,zhao3,Liu}.

\section{Conclusion and Discussion}
We present the correspondence relation between MI and fundamental excitations in the NLSE with both third-order and fourth-order effects.  Stable AD and WS$_{nr}$ soliton excitations are found to exist in MI regime outside of the MS circle (but not the MI regime inside of the MS circle), in sharp contrast to the correspondence relation for the simplest NLSE \cite{Zhaoling}, Hirota equation \cite{Liu}, and S-S equation \cite{zhao3}. We further test the evolution stability of the stable soliton numerically, which indicate they are robust against perturbations even in MI regime. The solitons with high energies in MI regime are caused by the fourth-order effects.

When the perturbation energy on CW background corresponding to AD or WS$_{nr}$ soliton tends to zero (the amplitude is small), the stable soliton evolution agree well with the stability property predicted by linear stability analysis. However, linear stability analysis fails to explain the stability of soliton with larger perturbation energy on the CW background, since their amplitude is large, for which the linear
stability analysis is not applicable. But rational WS obtained here and  \cite{zhao2,zhao3} suggest
that the linear stability analysis holds well for
large-amplitude perturbations just with the perturbation
energies tend to be zero. These characters indicate that the linear stability analysis does not hold for strong perturbations with nonzero perturbation energies. Therefore, it is still needed to develop some new ways to analyze the stability and dynamics of these type perturbations, and find out the underlying reasons for these striking characters. It is also meaningful to study excitation dynamics with other types fourth-order effects or higher-order effects to find out whether there are some new MI properties. The results here further indicate that different MI distributions would bring different excitation patterns.

\section*{Acknowledgments}
This work is supported by National Natural Science Foundation of
China (Contact No. 11475135), and the Education Department of Shaanxi Province, China (Contact No. 16JK1763).

\section*{Appendix: The generalized solution for fundamental nonlinear excitations\label{A}}
A generalized exact solution of the NLSE with both third-order and fourth-order effects is derived to describe types of fundamental nonlinear excitations. The solution is given as follows,
\begin{equation}
\label{equ:A1}
\psi_{1,2}=\left[1-\frac{8b\Gamma_{1,2}(\cosh(\phi+\delta_{1,2})+\cos(\varphi-\eta_{1,2}))}
{\Delta_{1,2}\cosh(\phi+\tau_{1,2})+\Xi_{1,2}\cos(\varphi-\rho_{1,2})}\right]Ae^{i\theta},
\end{equation}
where
\begin{eqnarray}
\label{equ:A2}
&&\phi=\zeta t-V_{1} z,~\varphi=\sigma t-V_{2} z,\nonumber\\
&&V_{1}=\zeta v_1-b \sigma v_2,~V_{2}=\sigma v_1+b \zeta v_2,\nonumber\\
&&v_1=\omega+\beta(2A^{2}+4b^{2}-3\omega^{2})+4\gamma(2A^{2}\omega+4b^{2}\omega-\omega^{3}),\nonumber\\
&&v_2=1-6\beta\omega+2\gamma(2A^{2}+4b^{2}-6\omega^{2}),\nonumber\\
&&\zeta=\left(\sqrt{\chi^2}+\chi\right)^{1/2}/\sqrt{2},\nonumber\\
&&\sigma=\pm\left(\sqrt{\chi^2}-\chi\right)^{1/2}/\sqrt{2},\nonumber\\
&&\chi=4b^2-4A^2.
\end{eqnarray}
The corresponding amplitude and phase notations are
\begin{eqnarray*}
&&\Gamma_{1}=\sqrt{\beta_1^2+\beta_2^2}, \Gamma_{2}=\sqrt{\alpha_1^2-\alpha_2^2}, \nonumber\\
&&\Delta_{1}=4A^{2}+\beta_{1}^{2}+\beta_{2}^{2},\nonumber\\
&&\Delta_{2}=\sqrt{(\alpha_3+\alpha_4)^2-16b^{2}\zeta^{2}},\nonumber\\
&&\Xi_{1}=-4A\beta_{1},\nonumber\\
&&\Xi_{2}=\sqrt{16b^{2}\sigma^{2}+(\alpha_3-\alpha_4)^2},\nonumber\\
\end{eqnarray*}
where $\cosh\delta_{1}=\frac{\beta_{1}}{\Gamma_{1}}$, $\sinh\delta_{1}=-\frac{\beta_{2}}{\Gamma_{1}}$, $\cosh\delta_{2}=\frac{\alpha_{1}}{\Gamma_{2}}$, $\sinh\delta_{2}=-\frac{\alpha_{2}}{\Gamma_{2}}$,
$\cos\eta_{1}=-\frac{4A^{2}+\beta_{1}^{2}+\beta_{2}^{2}}{4A\Gamma_{1}}$, $\sin\eta_{1}=-i\frac{4A^{2}-\beta_{1}^{2}-\beta_{2}^{2}}{4A\Gamma_{1}}$,
$\cos\eta_{2}=\frac{\alpha_{1}}{\Gamma_{2}}$, $\sin\eta_{2}=-i\frac{\alpha_{2}}{\Gamma_{2}}$,
$\tau_{1}=0$, $\cosh\tau_{2}=\frac{\alpha_{3}+\alpha_{4}}{\Delta_{2}}$, $\sinh\tau_{2}=-\frac{4b\zeta}{\Delta_{2}}$,
$\rho_{1}=0$, $\cos\rho_{2}=\frac{\alpha_{3}-\alpha{4}}{\Xi_{2}}$, $\sin\rho_{2}=\frac{4b\sigma}{\Xi_{2}}$  and $\alpha_1=2b,~\alpha_2=\zeta+i\sigma$,
$\alpha_3=4A^2+4b^2,~\alpha_4=\zeta^2+\sigma^2$, $\beta_{1}=2b+\zeta, \beta_{2}=\sigma$.

We find that the solution (\ref{equ:A1}) contain eight types fundamental nonlinear excitations, such as Kuznetsov-Ma breather, non-rational W-shaped soliton, anti-dark soliton, Akhmediev breather, periodic wave, W-shaped soliton train, rogue wave, and rational W-shaped soliton train. The existence conditions and explicit expressions of these fundamental nonlinear excitations are given in Table I.

\end{document}